\documentclass[11pt,twoside]{article}
\usepackage{graphicx}
\usepackage{subfigure}
\usepackage{fancyheadings}
\usepackage{amsmath}
\usepackage{amssymb}
\usepackage{cite}

\begin{document}
\newcommand{\pst}{\hspace*{1.5em}}

\newcommand{\rigmark}{\em Journal of Russian Laser Research}
\newcommand{\lemark}{\em Volume 35, Number 3, 2014}

\newcommand{\be}{\begin{equation}}
\newcommand{\ee}{\end{equation}}
\newcommand{\bm}{\boldmath}
\newcommand{\ds}{\displaystyle}
\newcommand{\bea}{\begin{eqnarray}}
\newcommand{\eea}{\end{eqnarray}}
\newcommand{\ba}{\begin{array}}
\newcommand{\ea}{\end{array}}
\newcommand{\arcsinh}{\mathop{\rm arcsinh}\nolimits}
\newcommand{\arctanh}{\mathop{\rm arctanh}\nolimits}
\newcommand{\bc}{\begin{center}}
\newcommand{\ec}{\end{center}}

\thispagestyle{plain}

\label{sh}

\begin{center} {\Large \bf
\begin{tabular}{c}
SEPARABILITY AND ENTANGLEMENT

\\[-1mm]
OF THE QUDIT X-STATE WITH $j=3/2$
\end{tabular}
 } \end{center}

\bigskip

\bigskip

\begin{center} {\bf
V.I. Man'ko$^{1}$ and L.A. Markovich$^{2*}$
}\end{center}

\medskip

\begin{center}
{\it
$^1$P.N. Lebedev Physical Institute, Russian Academy of Sciences\\
Leninskii Prospect 53, Moscow 119991, Russia

\smallskip

$^2$Institute of Control Sciences, Russian Academy of Sciences\\
Profsoyuznaya 65, Moscow 117997, Russia
}
\smallskip

$^*$Corresponding author e-mail:~~~kimo1~@~mail.ru\\
\end{center}

\begin{abstract}\noindent
The qudit state for $j=3/2$ with density matrix of the form corresponding to X-state of two-qubits is studied from the point of view of entanglement
and separability properties. The method of qubit portrait of qudit states is used to get the entropic inequalities for the entangled state of the single qudit.
The tomographic probability representation of the qudit X-state under consideration and its Shannon and q-entropic characteristics
are presented in explicit form.
\end{abstract}

\medskip


\section{Introduction}
It is known that the composite quantum system states, for example bipartite ones, are described by density operators $\widehat{\rho}(1,2)$ which provide the possibility to
construct reduced density operators $\widehat{\rho}(1)=Tr_2\widehat{\rho}(1,2)$ and $\widehat{\rho}(2)=Tr_1\widehat{\rho}(1,2)$ which describe the states of the subsystems 1 and 2, respectively. Recently it was observed \cite{Chernega,Chernega:14,Manko,OlgaMankoarxiv} that the quantum properties of systems without subsystems can be formulated by using invertible map of integers $1,2,3\ldots$ onto the pairs (triples, etc) of integers $(i,k)$, $j,k=1,2,\ldots$. Namely, the state density operator $\widehat{\rho}_{1}$ of system without subsystems, e.g. the state of single qudit $j=0,1/2,1,3/2,2,\ldots$ can be mapped onto density operator of the system containing the subsystems, (e.g. the state of two qudits) $\widehat{\rho}(1,2)$.
Thus, we have passibility to translate known properties of quantum correlations associated with structure of bipartite system like entanglement to the system without subsystems.
It is worthy to note that the quantum correlations, similar to correlations analogous to entanglement  for the single qudit were found and formulated in terms of quantum contextuality in \cite{Klichko}.
\par The aim of our work is to develop the remark presented in \cite{Chernega:14,OlgaMankoarxiv}
where the state of single qudit with $j=3/2$ was suggested to be studied in the aspect of entanglement existing in bipartite system  of two qubits.
\par We introduce $X$-state and construct for density matrix  $\rho_{3/2}$ of qudit with $j=3/2$ the analogs of qubit reduced density matrices $\rho(1)$ and $\rho(2)$,
respectively. We define the separability and entanglement properties of the density matrix $\rho_{3/2}$ using the invertible map $\rho_{3/2}\Leftrightarrow \rho(1,2)$.
All the density matrices are matrices of the operators, calculated in the corresponding bases. For example in the case of qudit with $j=3/2$ the basis is defined as set of vectors
$|3/2,m\rangle$, where $m=3/2,1/2,-1/2,-3/2$ and these vectors are eigenvectors of the operator $\widehat{J}_z$. The operator $\widehat{J}_z$ acting in four-dimensional
Hilbert space of the qudit state has the eigenvectors, i.e. $\widehat{J}_z|3/2,m\rangle=m|3/2,m\rangle$. In the case of two qubits the basis in four-dimensional
Hilbert space of the system states is defined as the set of vectors $|m_1m_2\rangle$, where $m_1,m_2=\pm1/2$.
These vectors are eigenvectors of two commuting operators $2\widehat{J}_{z_{1}}=\sigma_z\otimes1_2$ and  $2\widehat{J}_{z_{2}}=1_2\otimes\sigma_z$. The four-dimensional matrices of these two operators in the described basis have only diagonal nonzero matrix elements. The elements read $(2\widehat{J}_{z_{1}})_{1/2,1/2}=(2\widehat{J}_{z_{1}})_{1/2,-1/2}=
-(2\widehat{J}_{z_{1}})_{-1/2,1/2}=-(2\widehat{J}_{z_{1}})_{-1/2,-1/2}=(2\widehat{J}_{z_{2}})_{1/2,1/2}=(2\widehat{J}_{z_{2}})_{1/2,-1/2}=-(2\widehat{J}_{z_{2}})_{-1/2,1/2}=-(2\widehat{J}_{z_{2}})_{-1/2,-1/2}=1$. The matrix $\widehat{J}_{z}$ has the nonzero matrix elements in basis $|3/2,m\rangle$ as follows: $(\widehat{J}_{z})_{3/2,3/2}=3(\widehat{J}_{z})_{1/2,1/2}=-3(\widehat{J}_{z})_{-1/2,-1/2}=-(\widehat{J}_{z})_{-3/2,-3/2}=3/2$.
 The matrices $\widehat{J}_{z}$, $\widehat{J}_{z_{1}}$ and $\widehat{J}_{z_{2}}$ commute. In view of this, the corresponding observables can be measured simultaneously.
 We study the corresponding properties of the qudit and two-qubit systems in parallel. We define the separability and entanglement of the qudit   with $j=3/2$.
 We obtain the explicit formulas for von Neumann entropy and information and derive the entropic inequalities for $X$-states of the qudit with $j=3/2$.
\section{Qudit entropic inequality}
Let us define the density matrix of qudit state with spin $j=3/2$. This matrix has the form
\begin{eqnarray*}\rho_{3/2}&=&\left(
                                 \begin{array}{cccc}
                                   \rho_{3/2,3/2}& \rho_{3/2,1/2}& \rho_{3/2,-1/2}& \rho_{3/2,-3/2}\\
                                   \rho_{1/2,3/2}& \rho_{1/2,1/2}& \rho_{1/2,-1/2}& \rho_{1/2,-3/2}\\
                                   \rho_{-1/2,3/2}& \rho_{-1/2,1/2}& \rho_{-1/2,-1/2}& \rho_{-1/2,-3/2}\\
                                   \rho_{-3/2,3/2}& \rho_{-3/2,1/2}& \rho_{-3/2,-1/2}& \rho_{-3/2,-3/2}\\
                                 \end{array}
                               \right)
                               \end{eqnarray*}
and has the properties $\rho_{3/2}=\rho_{3/2}^{\dagger}$, $Tr\rho_{3/2}=1$. The eigenvalues of the matrix are nonnegative.
\par Let us introduce an invertible map of indices $1\leftrightarrow 3/2$, $2\leftrightarrow1/2$, $3\leftrightarrow-1/2$, $4\leftrightarrow-3/2$. Applying this mapping to the density matrix $\rho_{3/2}$ one can write
\begin{eqnarray}\label{1}\rho_{3/2}&=&\left(
                                 \begin{array}{cccc}
                                   \rho_{11}& \rho_{12}& \rho_{13}& \rho_{14}\\
                                   \rho_{21}& \rho_{22}& \rho_{23}& \rho_{24}\\
                                   \rho_{31}& \rho_{32}& \rho_{33}& \rho_{34}\\
                                   \rho_{41}& \rho_{42}& \rho_{43}& \rho_{44}\\
                                 \end{array}
                               \right).
                               \end{eqnarray}
If $\rho_{12}=\rho_{13}=\rho_{21}=\rho_{31}=\rho_{24}=\rho_{34}=\rho_{42}=\rho_{43}=0$ then the density matrix \eqref{1} has the view of two-qubit $X$-state density matrix
                               \begin{eqnarray}\label{2}
\rho_{3/2}^{X}&=&\left(
                     \begin{array}{cccc}
                       \rho_{11} & 0 & 0 & \rho_{14}\\
                       0 & \rho_{22}& \rho_{23} & 0 \\
                       0 & \rho_{32} & \rho_{33} & 0 \\
                       \rho_{41} & 0 & 0 & \rho_{44} \\
                     \end{array}
                   \right)=\left(
                     \begin{array}{cccc}
                       \rho_{11} & 0 & 0 & \rho_{14}\\
                       0 & \rho_{22}& \rho_{23} & 0 \\
                       0 & \rho_{23}^{\ast} & \rho_{33} & 0 \\
                      \rho_{14}^{\ast} & 0 & 0 & \rho_{44} \\
                     \end{array}
                   \right),
\end{eqnarray}
where $\rho_{11},\rho_{22},\rho_{33},\rho_{44}$ are real positive and $\rho_{23},\rho_{14}$ complex quantities. In fact if we will apply the map
$1\leftrightarrow 1/2,1/2$, $2\leftrightarrow1/2,-1/2$, $3\leftrightarrow-1/2,1/2$, $4\leftrightarrow-1/2,-1/2$ to the matrix elements of \eqref{2} the new matrix will have view of two-qubit density matrix.
The latter matrix has the unit trace and is nonnegative if $\rho_{22}\rho_{33}\geq|\rho_{23}|^2$, $\rho_{11}\rho_{44}\geq|\rho_{14}|^2$.
The eigenvalues of \eqref{2} are (see for example \cite{Mazhar})
\begin{eqnarray*}\lambda_1 &=&\frac{1}{2}\left(\rho_{11}+\rho_{44}+\sqrt{\left(\rho_{11}-\rho_{44}\right)^2+4|\rho_{14}|^2}\right),\\
\lambda_2 &=&\frac{1}{2}\left(\rho_{11}+\rho_{44}-\sqrt{\left(\rho_{11}-\rho_{44}\right)^2+4|\rho_{14}|^2}\right),\\
\lambda_3 &=&\frac{1}{2}\left(\rho_{22}+\rho_{33}+\sqrt{\left(\rho_{22}-\rho_{33}\right)^2+4|\rho_{23}|^2}\right),\\
\lambda_4&=&\frac{1}{2}\left(\rho_{22}+\rho_{33}-\sqrt{\left(\rho_{22}-\rho_{33}\right)^2+4|\rho_{23}|^2}\right).
\end{eqnarray*}
The reduced density matrices are defined as
\begin{eqnarray*}\rho(1)&=&\left(
                              \begin{array}{cc}
                                \rho_{11}+\rho_{22} & \rho_{13}+\rho_{24} \\
                               \rho_{31}+\rho_{42} &  \rho_{33}+\rho_{44} \\
                              \end{array}
                            \right)
=\left(
                            \begin{array}{cc}
                              \rho_{11}+\rho_{22} & 0 \\
                              0 & \rho_{33}+\rho_{44} \\
                            \end{array}
                          \right),\\
                          \rho(2)&=&\left(
                              \begin{array}{cc}
                                \rho_{11}+\rho_{33} & \rho_{12}+\rho_{34} \\
                               \rho_{31}+\rho_{42} &  \rho_{22}+\rho_{44} \\
                              \end{array}
                            \right)=\left(
                            \begin{array}{cc}
                              \rho_{11}+\rho_{33}& 0 \\
                              0 & \rho_{22}+\rho_{44} \\
                            \end{array}
                          \right).
\end{eqnarray*}
Hence the von Neumann entropies can be obtained as
\begin{eqnarray}\label{4}S_1 &=& -Tr\rho(1)\ln\rho(1) = - Tr(\rho_{11}+\rho_{22})\ln(\rho_{11}+\rho_{22})-Tr(\rho_{33}+\rho_{44})\ln(\rho_{33}+\rho_{44}),\\
S_2 &=&-Tr\rho(2)\ln\rho(2)= - Tr(\rho_{11}+\rho_{33})\ln(\rho_{11}+\rho_{33})-Tr(\rho_{22}+\rho_{44})\ln(\rho_{22}+\rho_{44}),\\ \nonumber
S_{12} &=&-Tr\rho_{3/2}^{X}\ln\rho_{3/2}^{X}=-\sum\limits_{i=1}^{4}\lambda_i\ln\lambda_i
\end{eqnarray}
and the quantum information is defined as
\begin{eqnarray}\label{12}I_q^{X}&=&S_1 +S_2-S_{12}.
\end{eqnarray}
Obviously, the quantum information satisfies the inequality $I_q\geq 0$. Hence
\begin{eqnarray*}&&- Tr(\rho_{11}+\rho_{22})\ln(\rho_{11}+\rho_{22})-Tr(\rho_{33}+\rho_{44})\ln(\rho_{33}+\rho_{44})\\
&-&Tr(\rho_{11}+\rho_{33})\ln(\rho_{11}+\rho_{33})-Tr(\rho_{22}+\rho_{44})\ln(\rho_{22}+\rho_{44})+\sum\limits_{i=1}^{4}\lambda_i\ln\lambda_i\geq 0.
\end{eqnarray*}
The subadditivity condition holds for the qudit system with $j=3/2$ and this single qudit system does not contain subsystems.
\section{Negativity and concurrence}
Using the map of density matrix of two-qubit system onto density matrix for state of qudit with $j=3/2$ we define analog of positive partial transpose operation known for bipartite
system states to the single-qudit state.  We get the matrix
                               \begin{eqnarray*}
\rho_{3/2}^{Xppt}&=&\left(
                     \begin{array}{cccc}
                       \rho_{11} & 0 & 0 & \rho_{23}\\
                       0 & \rho_{22}& \rho_{14} & 0 \\
                       0 & \rho_{14}^{\ast} & \rho_{33} & 0 \\
                      \rho_{23}^{\ast} & 0 & 0 & \rho_{44} \\
                     \end{array}
                   \right),
\end{eqnarray*}
with eigenvalues
\begin{eqnarray*}\lambda_1^{ppt} &=&\frac{1}{2}\left(\rho_{11}+\rho_{44}+\sqrt{\left(\rho_{11}-\rho_{44}\right)^2+4|\rho_{23}|^2}\right),\\
\lambda_2^{ppt} &=&\frac{1}{2}\left(\rho_{11}+\rho_{44}-\sqrt{\left(\rho_{11}-\rho_{44}\right)^2+4|\rho_{23}|^2}\right),\\
\lambda_3^{ppt} &=&\frac{1}{2}\left(\rho_{22}+\rho_{33}+\sqrt{\left(\rho_{22}-\rho_{33}\right)^2+4|\rho_{14}|^2}\right),\\
\lambda_4^{ppt}&=&\frac{1}{2}\left(\rho_{22}+\rho_{33}-\sqrt{\left(\rho_{22}-\rho_{33}\right)^2+4|\rho_{14}|^2}\right).
\end{eqnarray*}
In the case where the following inequality holds
\begin{eqnarray*}&&|\lambda_1^{ppt}|+|\lambda_2^{ppt}|+|\lambda_3^{ppt}|+|\lambda_4^{ppt}|>1
\end{eqnarray*}
we will interpret the sum in the left hand side of this inequality as negativity parameter characterizing the state of qudit with $j=3/2$. We interpret the qudit state with matrix
\eqref{2} satisfying  the inequality as entangled state.
It is known that $X$-state of two qubits is entangled if either $\rho_{22}\rho_{33}<|\rho_{14}|^2$ or $\rho_{11}\rho_{44}<|\rho_{23}|^2$. Both conditions cant be fulfilled simultaneously \cite{Mazhar}.
\par The concurrence is defined as
 \begin{eqnarray}\label{C}C(\rho)= \max\{0,\sqrt{\lambda_1}-\sqrt{\lambda_2}-\sqrt{\lambda_3}-\sqrt{\lambda_4}\},
 \end{eqnarray}
 where $\lambda_i,i=1,2,3,4$ are the square-roots of the eigenvalues of matrix $\rho_{3/2}^{X}\widetilde{\rho}_{3/2}^{X}$ in decreasing order.  We use the definition of concurrence which
 is analogues to concurrence for two-qubit system.  The matrix $\widetilde{\rho}_{3/2}^{X}$
 is obtained by spin flip operation on the qudit density matrix \eqref{2}
                               \begin{eqnarray}\label{3}
\widetilde{\rho}_{3/2}^{X}&=&(\sigma_y\otimes\sigma_y)\rho_{3/2}^{X\ast}(\sigma_y\otimes\sigma_y),
\end{eqnarray}
where $\sigma_y$ is a Pauli matrix and
\begin{eqnarray*}(\sigma_y\otimes\sigma_y)&=&\left(
                     \begin{array}{cccc}
                       0 & 0 & 0 &-1\\
                       0 &0& 1 & 0 \\
                       0 & 1& 0 & 0 \\
                      -1 & 0 & 0 &0 \\
                     \end{array}
                   \right).
\end{eqnarray*}
Hence \eqref{3} is
\begin{eqnarray}\label{5}
\widetilde{\rho}_{3/2}^{X}&=&\left(
                     \begin{array}{cccc}
                       \rho_{44} & 0 & 0 & \rho_{14}\\
                       0 & \rho_{33}& \rho_{23} & 0 \\
                       0 & \rho_{23}^{\ast} & \rho_{22} & 0 \\
                      \rho_{14}^{\ast} & 0 & 0 & \rho_{11} \\
                     \end{array}
                   \right).
\end{eqnarray}
Multiplying matrices \eqref{2} and \eqref{5} we get
\begin{eqnarray*}
\rho_{3/2}^{X}\widetilde{\rho}_{3/2}^{X}&=&\left(
                     \begin{array}{cccc}
                       |\rho_{14}|^2+\rho_{11}\rho_{44} & 0 & 0 & 2\rho_{11}\rho_{14}\\
                       0 & |\rho_{23}|^2+\rho_{22}\rho_{33}& 2\rho_{22}\rho_{23} & 0 \\
                       0 & 2\rho_{33}\rho_{23}^{\ast} & |\rho_{23}|^2+\rho_{22}\rho_{33} & 0 \\
                      2\rho_{44}\rho_{14}^{\ast} & 0 & 0 & |\rho_{14}|^2+\rho_{11}\rho_{44} \\
                     \end{array}
                   \right).
\end{eqnarray*}
The eigenvalues of the obtained matrix are
\begin{eqnarray*}\lambda_1^{C} &=&|\rho_{14}|^2+\rho_{11}\rho_{44}-2\sqrt{\rho_{11}\rho_{44}|\rho_{14}|^2}=(|\rho_{14}|-\sqrt{\rho_{11}\rho_{44}})^2,\\
\lambda_2^{C} &=&|\rho_{14}|^2+\rho_{11}\rho_{44}+2\sqrt{\rho_{11}\rho_{44}|\rho_{14}|^2}=(|\rho_{14}|+\sqrt{\rho_{11}\rho_{44}})^2,\\
\lambda_3^{C} &=&|\rho_{23}|^2+\rho_{22}\rho_{33}-2\sqrt{\rho_{22}\rho_{33}|\rho_{23}|^2}=(|\rho_{23}|-\sqrt{\rho_{22}\rho_{33}})^2,\\
\lambda_4^{C}&=&|\rho_{23}|^2+\rho_{22}\rho_{33}+2\sqrt{\rho_{22}\rho_{33}|\rho_{23}|^2}=(|\rho_{23}|+\sqrt{\rho_{22}\rho_{33}})^2.
\end{eqnarray*}
Hence, it can be deduced that the concurrence \eqref{C} is determined by (see for example \cite{Hedemann})
 \begin{eqnarray*}C(\rho)= \max\{0,2|\rho_{23}|-2\sqrt{\rho_{11}\rho_{44}},2|\rho_{14}|-2\sqrt{\rho_{22}\rho_{33}}\}.
 \end{eqnarray*}
 \section{The Werner state for spin $j=3/2$ with two parameters}
\begin{figure}[ht]
\begin{center}
\begin{minipage}[ht]{0.49\linewidth}
\includegraphics[width=1\linewidth]{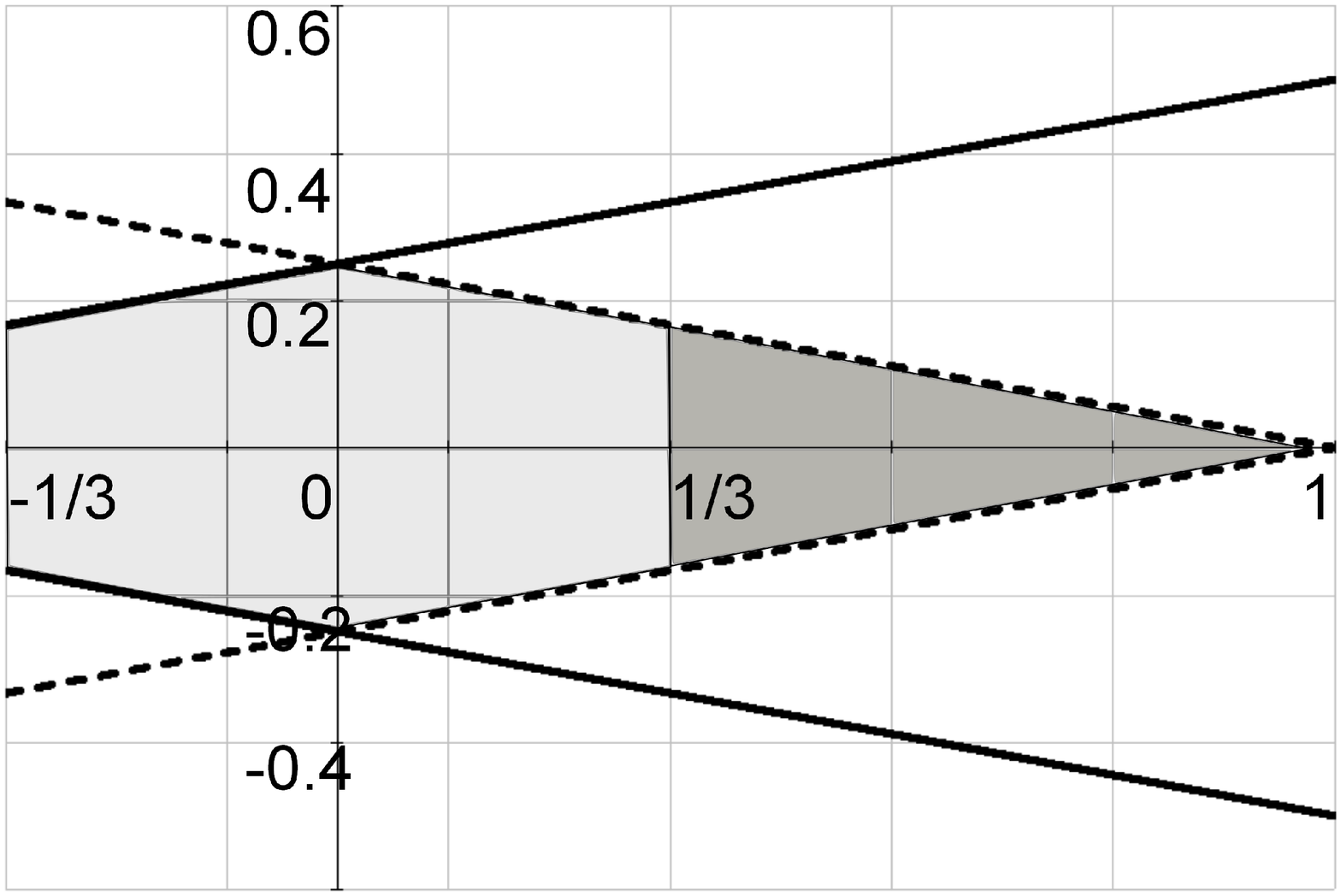}
\vspace{-4mm}
\caption{The gray area shows the area of parameters $p$ ($X$-axis), $b$ ($Y$-axis) for which the matrix $\rho^{Wppt}_{3/2}(p,b)$ is nonnegative}
\label{fig:1}
\end{minipage}
\hfill
\begin{minipage}[ht]{0.49\linewidth}
\includegraphics[width=1\linewidth]{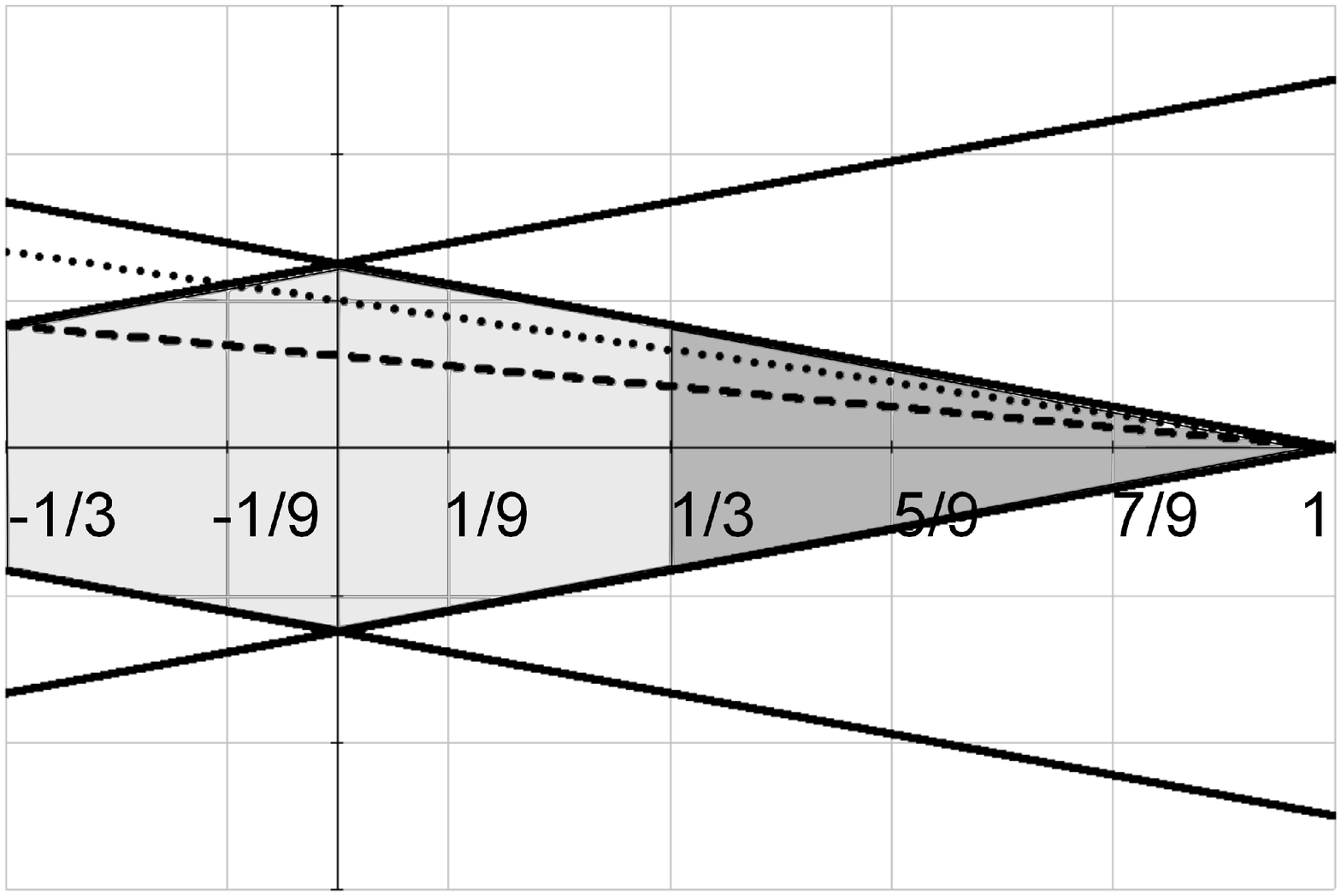}
\vspace{-4mm}
\caption{$b=\frac{1-p}{8}$ (dashed line) is full in gray area, $b=\frac{1-p}{5}$ (dotted line) is partly in gray area}
\label{fig:2}
\end{minipage}
\end{center}
\end{figure}

As en example of $X$-state density matrix of qudit state with spin $j=3/2$ can be taken the following two-parametric matrix
\begin{eqnarray}\label{10}\rho^{W}_{3/2}(p,b)&=&\left(
                     \begin{array}{cccc}
                       \frac{1+p}{4} & 0 & 0 & \frac{p}{2}\\
                       0 & \frac{1-p}{4}& b & 0 \\
                       0 & b & \frac{1-p}{4} & 0 \\
                       \frac{p}{2} & 0 & 0 & \frac{1+p}{4} \\
                     \end{array}
                   \right).
\end{eqnarray}
To provide the positivity of \eqref{10} the parameters are $-\frac{1}{3}\leq p\leq1$, $\frac{1-p}{4}\geq |b|$.
The eigenvalues of \eqref{10} are the following
\begin{eqnarray*}&&\lambda_1^W = \frac{1+3p}{4}, \quad \lambda_{2}^W = \frac{1-p}{4}, \quad \lambda_{3}^W = \frac{1-p}{4}-b, , \quad \lambda_{4}^W = \frac{1-p}{4}+b.
\end{eqnarray*}
After ppt the matrix \eqref{10} is
\begin{eqnarray}\label{11}\rho^{Wppt}_{3/2}(p,b)&=&\left(
                     \begin{array}{cccc}
                       \frac{1+p}{4} & 0 & 0 &b \\
                       0 & \frac{1-p}{4}& \frac{p}{2} & 0 \\
                       0 & \frac{p}{2} & \frac{1-p}{4} & 0 \\
                       b & 0 & 0 & \frac{1+p}{4} \\
                     \end{array}
                   \right).
\end{eqnarray}
To provide the positivity of \eqref{11} is must be $\frac{1+p}{4}\geq |b|$.
The area where matrix \eqref{11} is nonnegative is shown in Figure \ref{fig:1}. With solid lines are shown lines $b=\frac{1+p}{4}$, $b=-\frac{1+p}{4}$, with dashed $b=\frac{1-p}{4}$, $b=\frac{1-p}{4}$. Intersection of the domains $\frac{1-p}{4}\geq |b|$ and $\frac{1+p}{4}\geq |b|$ are shown as gray area.  Dark gray triangle is the area where $\frac{1}{3}\leq p\leq1$ (entanglement state).
\par Thus, the parameter $b$ must be entirely fall into a gray area. In Figure \ref{fig:2} are shown two kinds of parameters $b$. With the dashed line $b=\frac{1-p}{8}$
which falls entirely in the area where the density matrix \eqref{11} is nonnegative. The parameter $b=\frac{1-p}{5}$ goes beyond this area at the point $p=-\frac{1}{9}$ that is shown with the dotted line.
\par
\begin{figure}[ht]
\begin{center}
\begin{minipage}[ht]{0.49\linewidth}
\includegraphics[width=1\linewidth]{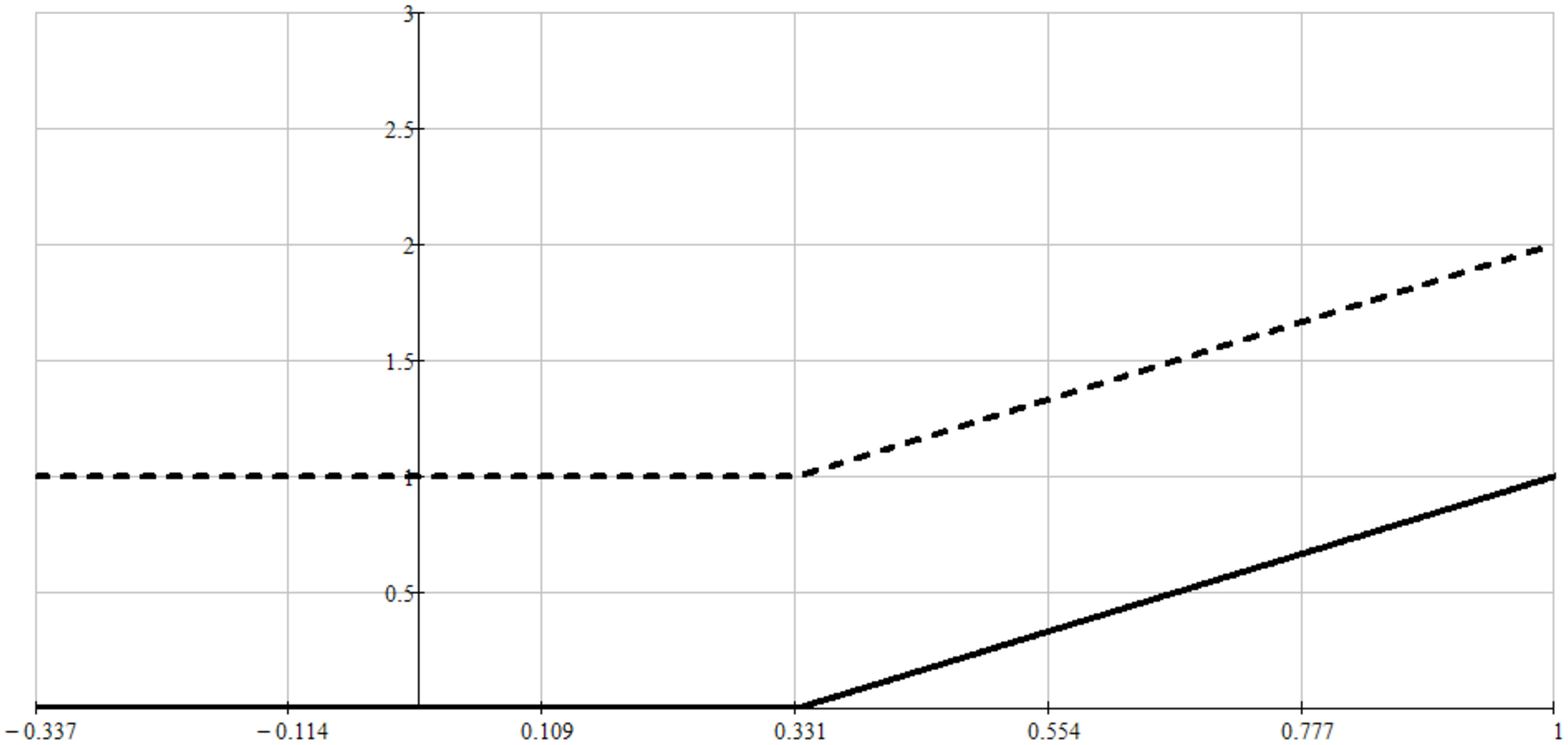}
\vspace{-4mm}
\caption{Negativity (dashed line) and concurrence (solid line) for $b=\frac{1-p}{8}$}
\label{fig:3}
\end{minipage}
\hfill
\begin{minipage}[ht]{0.49\linewidth}
\includegraphics[width=1\linewidth]{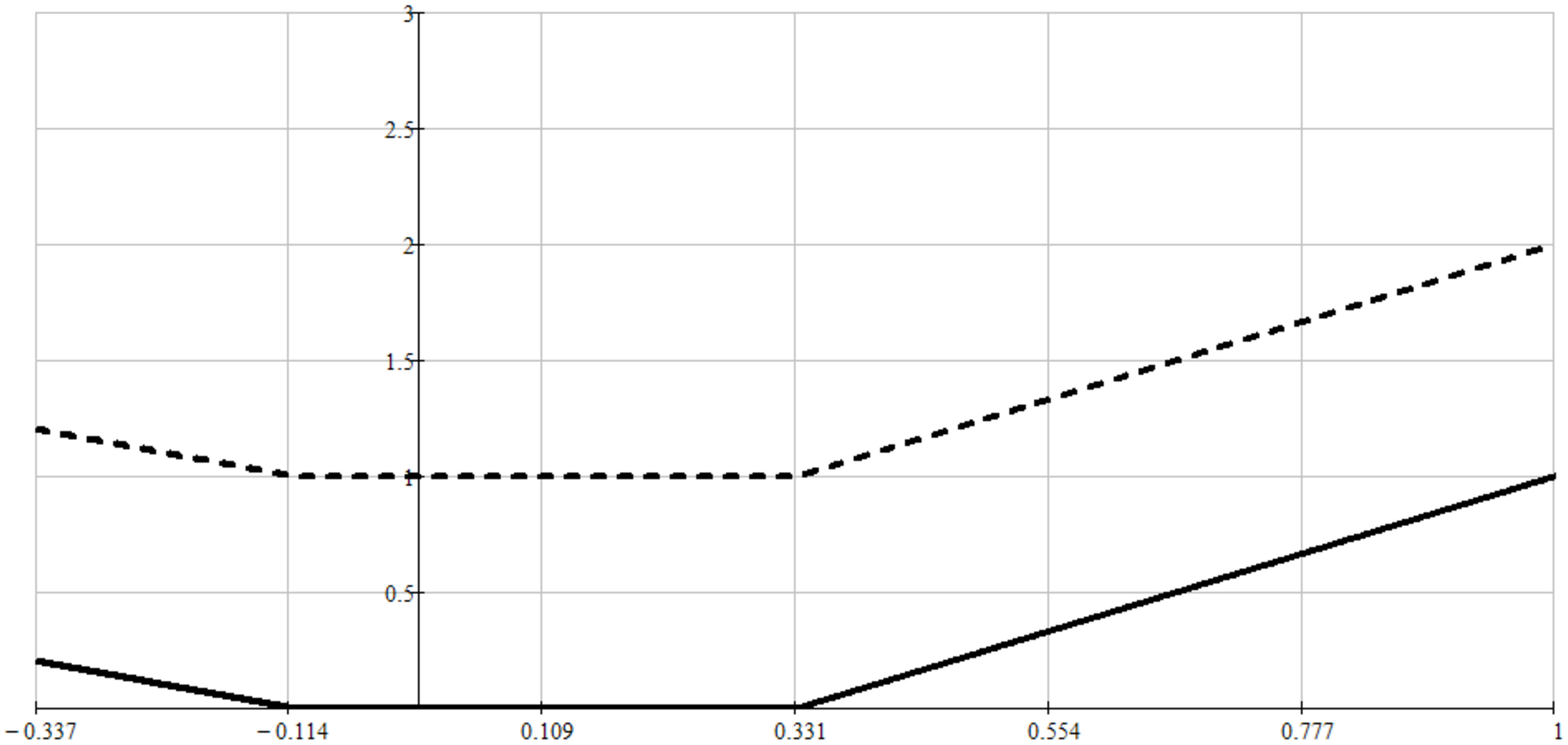}
\vspace{-4mm}
\caption{Negativity (dashed line) and concurrence (solid line) for $b=\frac{1-p}{5}$}
\label{fig:4}
\end{minipage}
\end{center}
\end{figure}

 The eigenvalues of \eqref{11} are the following
\begin{eqnarray*}&&\lambda_1^{Wppt} = \frac{1-3p}{4}, \quad \lambda_{2}^{Wppt}= \frac{1+p}{4}, \quad \lambda_{3}^{Wppt}= \frac{1+p}{4}-b, , \quad \lambda_{4}^{Wppt} = \frac{1+p}{4}+b.
\end{eqnarray*}
The negativity is defined as
\begin{eqnarray}\label{Neg}&&\left|\frac{1-3p}{4}\right|+\left|\frac{1+p}{4}\right|+\left|\frac{1+p}{4}-b\right|+\left|\frac{1+p}{4}+b\right|>1
\end{eqnarray}
and concurrence is
 \begin{eqnarray}\label{Con}C(\rho)= \max\left\{0,2|b|-2\frac{1+p}{4},2\left|\frac{p}{2}\right|-2\frac{1-p}{4}\right\}.
 \end{eqnarray}
The \eqref{Neg} and \eqref{Con}  are shown in Figure \ref{fig:3} for parameter  $b=\frac{1-p}{8}$ and looks completely like for two-partite system concurrence and negativity. For parameter  $b=\frac{1-p}{5}$ that is not completely in area where \eqref{11} is nonnegative the negativity and concurrence are shown In Figure \ref{fig:4}. In area $-\frac{1}{3}\leq p\leq-\frac{1}{9}$ the density matrix is not nonnegative and \eqref{Neg} and \eqref{Con} are not valid.
\par There are Tsallis \cite{Tsallis} and Renyi \cite{Renyi} entropies depending on extra parameter q and they  are called $q$-entropies.
The classical $q$-entropies for $\overrightarrow{p}=(p_1=\rho_{11},p_2=\rho_{22},p_3=\rho_{33},p_4=\rho_{44})$ are
 \begin{eqnarray*}T_{q}^{T}&=&\frac{1}{1-q}\left(\sum\limits_{i=1}^{4}p_i^q-1\right),\quad
 T_q^{R} =\frac{1}{1-q}\ln\left(\sum\limits_{i=1}^{4}p_i^q\right).
\end{eqnarray*}
The introduced $q$-entropies can be considered as entropy of "two-qubit" composite system.
In view of this the entropy of the qudit with $j=3/2$ has to satisfy the relations which are known for the state 
of the two-qubit system.
We study these inequalities in future publication.
\section{Conclusion}
To conclude we point out the main results of our work which present the extension of the approach used in \cite{Man,Lub}.
We introduced and studied the $X$-state of qudit with $j=3/2$. We obtained the entropic inequalities for the single qudit with $j=3/2$
which are analogs of the inequalities for the composite system of two qubits.
We developed approach of \cite{Chernega,Chernega:14,OlgaMankoarxiv,Manko}  and used notion of separability and entanglement 
of single qudit state which is the state of a system without subsystems. The concurrence and negativity were introduced in our work 
to characterize the degree of entanglement of the single qudit with $j=3/2$.
\section*{Acknowledgments}
\pst
L. A. M. acknowledges the financial support provided within the Russian Foundation for Basic Research, grant 13-08-00744 A.


\begin{thebibliography}{99}
\bibitem{Chernega}V.N. Chernega, O.V. Man'ko, V.I. Man'ko, {\it Generalized qubit portrait of the qutritstate density matrix}, {\sl J. Russ. Laser Res.}, \textbf{34 (4)}, 383--387 (2013).
\bibitem{Chernega:14}V.N. Chernega, O.V. Man'ko, {\it Tomographic and improved subadditivity conditions for two qubits and qudit with j = 3/2}, {\sl J. Russ. Laser Res.}, \textbf{35 (1)}, 27--38 (2014).
\bibitem{OlgaMankoarxiv}V.N. Chernega, O.V. Man'ko, V.I. Man'ko, {\it Subadditivity condition for spin-tomograms and density matrices of arbitrary composite and noncomposite qudit systems}, {\sl http://arxiv.org/abs/1403.2233}, (2014).
\bibitem{Manko}M.A. Man'ko and  V.I. Man'ko, arXiv:1312.6988 (2013), to appear in Physica Scripta (2014).
\bibitem{Klichko} A.A. Klyachko, M.A. Can, S. Binicioglu  and  A.S. Shumovsky, {\it Simple test for hidden variables in spin-1 systems}, {\sl Phys. Rev. Lett.}, \textbf{101}, 020403 (2008).
    \bibitem{Mazhar}A. Mazhar, A. R. P. Rau, G. Alber, {\it Quantum discord for two-qubit X-states}, {\sl Phys. Rev. A}, \textbf{81}, 042105 (2010).
\bibitem{Hedemann}S. R. Hedemann, {\it Evidence that All States Are Unitarily Equivalent to X States of the Same Entanglement}, {\sl http://arxiv.org/abs/1310.7038}, (2014)
\bibitem{Tsallis}C. Tsallis, {\it}, {\sl J. Stat. Phys.}, \textbf{52}, 479 (1988).
\bibitem{Renyi}A. R\'{e}nyi, {\it Probability Theory}, {\sl North-Holland, Amsterdam}, (1970).

\bibitem{Man} V.I. Man'ko and  L.A. Markovich, {\it Entropic inequalities and properties of some special functions},  {\sl J. Russ. Laser Res.}, \textbf{35 (2)}, 200--210 (2014)
\bibitem{Lub}V.I. Man'ko and L.A. Markovich, {\it New inequalities for quantum  von Neumann and tomographic mutual information},  {\sl J. Russ. Laser Res.}, \textbf{35(3)},(2014)
\end{thebibliography}
\end{document}